\documentstyle[amssymb,aps,12pt]{revtex}

\begin{document}
\draft
\author{S. N. Dolya$^{1}$ and O. B. Zaslavskii$^{2}$}
\address{$^{1}$B. Verkin Institute for Low Temperature Physics and\\
Engineering,\\
47 Lenin Prospekt, Kharkov 61103, Ukraine\\
E-mail: dolya@ilt.kharkov.ua\\
$^{2}$Department of Mechanics and Mathematics, Kharkov V.N. Karazin's\\
National\\
University, Svoboda\\
Sq.4, Kharkov 61077, Ukraine\\
E-mail: aptm@kharkov.ua}
\title{Quasi-exactly solvable quartic Bose Hamiltonians}
\maketitle

\begin{abstract}
We consider Hamiltonians, which are even polynomials of the forth order with
the respect to Bose operators. We find subspaces, preserved by the action of
Hamiltonian These subspaces, being finite-dimensional, include, nonetheless,
states with an {\it infinite} number of quasi-particles, corresponding to
the original Bose operators. The basis functions look rather simple in the
coherent state representation and are expressed in terms of the degenerate
hypergeometric function with respect to the complex variable labeling the
representation. In some particular degenerate cases they turn (up to the
power factor) into the trigonometric or hyperbolic functions, Bessel
functions or combinations of the exponent and Hermite polynomials. We find
explicitly the relationship between coefficients at different powers of Bose
operators that ensure quasi-exact solvability of Hamiltonian.
\end{abstract}



\section{Introduction}

The conception of quasi-exactly solvable (QES) systems, discovered in 1980s 
\cite{raz1} - \cite{book}, received in recent years much attention both from
viewpoint of physical applications and their inner mathematical beauty. In
turned out that in quantum mechanics there exists a peculiar class of
objects that occupy an intermediate place between exactly solvable and
non-solvable models in the sense that in an infinite Hilbert space a finite
part of a spectrum is singled out within which eigenvectors and eigenvalues
can be found from an algebraic equation of a finite degree - in other words,
a partial algebraization of the spectrum occurs. For one-dimensional QES
models corresponding QES Hamiltonians possess hidden group structure based
on $sl(2,R)$ algebra. Thus, they have direct physical meaning, being related
to quantum spin systems \cite{uz}.

Meanwhile, the notion of QES systems is not constrained by potential models
and can have nontrivial meaning for any kind of infinite-dimensional
systems. In the first place, it concerns Bose Hamiltonians whose physical
importance is beyond doubt. Here one should distinguish two cases. First, it
turns out that some systems of two interacting particles or quasi-particles
with Bose operators of creation and annihilation $a$, $a^{+}$ and $b$, $%
b^{+} $can be mapped on the problem for a particle moving in a certain type
of one-dimensional potentials and, remarkably, these potentials belong just
to the QES type \cite{sb} - \cite{deb}. In particular, such a type of
Hamiltonians is widely spread in quantum optics and physics of magnetism 
\cite{uz}. The aforementioned mapping works only for a special class of Bose
Hamiltonians which possess an integral of motion. Then the procedure is
performed in three steps: (i) all Hilbert space splits in a natural way to
different pieces with respect to the values of an integral of motion, (ii)
in each piece the Schr\"{o}dinger equation takes a finite-difference form,
(iii) it is transformed into the differential equation by means of
introducing a generating function. In so doing, the integral of motion under
discussion represents a linear combination of numbers of particles $a^{+}a$
and $b^{+}b$.

The second kind of Bose systems looks much more usual - it is simply some
polynomial with respect to Bose operators of creation and annihilation of
one particle. The fact that only one pair $a$, $a^{+}$ enters Hamiltonian,
deprives us, by contrast with the first case, of the possibility to
construct a simple integral of motion - in this sense the eigenvalue problem
becomes more complicated. In general, the solutions of the Shr\"{o}dinger
equation contain infinite numbers of quasi-particles and only approximate or
numerical methods can be applied to such systems. However, as was shown
recently \cite{bose}, if the coefficients at different powers of $a$, $a^{+}$
are selected in a proper way, in some cases a {\it finite-dimensional}
closed subspace is singled out and algebraization of the spectrum occurs
similar to what happens in ''usual'' QES potential models or differential
equations. In so doing, the eigenvectors belonging to the subspace under
discussion, can be expressed as a finite linear combination of eigenvectors
of an harmonic oscillator and, thus, contain a finite number of
quasi-particles \cite{bose}.

In the present article we extend the approach of \cite{bose} and consider
much more general classes of Hamiltonians. Their distinctive feature
consists in that the relevant basis functions that compose a
finite-dimensional subspace, look very much unlike the wave functions of a
harmonic oscillator. As a result, we obtain QES models with an {\it infinite}
numbers of quasi-particles in this {\it finite}-dimensional subspace.
Bearing in mind physical application, we make emphasis on Hermitian
Hamiltonians, although our approach is applicable to more general QES Bose
operators without demand of Hermiticity.

\section{Bose Hamiltonians as differential operators and structure of
invariant subspaces}

Consider the operator which is the even polynomial of the forth degree with
respect to Bose operators of creation $a^{+}$ and annihilation $a$. It can
be written in the form 
\begin{equation}
H=a_{++}K_{+}^{2}+a_{--}K_{-}^{2}+a_{00}K_{0}^{2}+a_{0-}K_{0}K_{-}+a_{+0}K_{+}K_{0}+a_{0}K_{0}+a_{-}K_{-}+a_{+}K_{+}%
\text{,}  \label{hk}
\end{equation}
where

\begin{equation}
K_{0}=\frac{1}{2}\left( a^{+}a+\frac{1}{2}\right) ,K_{-}=\frac{a^{2}}{2}%
,K_{+}=\frac{a^{+2}}{2},  \label{ki}
\end{equation}

\begin{equation}
\lbrack K_{0},\text{ }K_{\pm }]=\pm K_{\pm }\text{, }[K_{+},K_{-}]=-2K_{0}%
\text{.}  \nonumber
\end{equation}

The Casimir operator $C=K_{0}^{2}-\frac{1}{2}(K_{+}K_{-}+K_{-}K_{+})\equiv -%
\frac{3}{16}$.

We will use the coherent state representation in which 
\begin{equation}
a\rightarrow \frac{\partial }{\partial z}\text{, }a^{+}\rightarrow z
\label{az}
\end{equation}
After substitution into (\ref{hk}) Hamiltonian $H(a^{+},a)$ becomes a
differential operator $H(z,\frac{\partial }{\partial z})$. In the previous
article \cite{bose} we discussed Bose systems that possess the invariant
subspace of the form $F=span\{z^{n}\}\,$or $span\{z^{2n}\}$. The first
natural step towards generalization consists in considering subspaces (with $%
N$ fixed ) 
\begin{equation}
F=span\{u_{n}\}\text{, }u_{n}=z^{2n}u\text{, }n=0,1,2...N\text{,}  \label{f1}
\end{equation}
for which the following procedure should be realized. (i) The action of
operators of $K_{i}$ on the functions $u_{n}$ should lead to the linear
combinations of functions from the same set $\{u_{n}\}$, (ii) by the
selection of appropriate coefficients in (\ref{hk}), we achieve the subspace 
$F$ to be closed under the action of Hamiltonian $H$. We would like to
stress that the condition (i) does not forbid $u_{n}$ with $n>N$ to appear
in terms like $K_{i}u_{n}$ but the condition (ii) rules out such functions
from $Hu_{n}$ (recall that we consider Hamiltonians which are
quadratic-linear combinations of $K_{i}$).

It is seen from (\ref{ki}), (\ref{az}) that the operators $K_{i}$ contain $z$
and $\frac{\partial }{\partial z}$. Therefore, it is convenient to assume
that differentiation of $u(z)$ gives rise to $u$ up to the factor that
contains powers of $z$. The corresponding choice is not unique. In the
present article we restrict ourselves to one of the simplest possibilities
that leads to nontrivial solutions. To this end, we choose $u$ that obeys
the differential equation 
\begin{equation}
u^{\prime }=A(z)u\text{, }A(z)=(\frac{\beta }{z}+2\rho z)\text{.}  \label{ud}
\end{equation}
We will show below that the choice (\ref{ud}) relates $K_{i}u_{n}$ to $u_{n}$%
, $u_{n\pm 1}$ that, in turn, allows us to formulate the conditions of cut
off for Hamiltonian in the form of algebraic equations which its
coefficients obey. It follows from (\ref{ud}) that $u=z^{\beta }\exp (\rho
z^{2})$.{\large \ }To ensure asymptotic analytic behavior near $z=0$, we
demand that $\beta =0,1,2...$ Now let us take into account some basic
properties of coherent states (see, e.g. Ch. 7 of Ref. \cite{coh}). Our
functions $u_{n}(z)$ must belong to the Bargmann-Fock space. It means that
they should obey the conditions of integrability and analyticity. The
condition of integrability for any two functions $f$, $g$ from our space 
\begin{equation}
\int dzdz^{*}f^{*}ge^{-zz^{*}}<\infty \text{,}  \label{conv}
\end{equation}
entails, for our choice of $u$, $\left| \rho \right| <1/2$.{\large \ }

Taking into account eq. (\ref{ud}), it is straightforward to show that 
\begin{equation}
K_{+}u_{n}=C_{+}u_{n+1}\text{,}  \label{c1}
\end{equation}
\begin{equation}
K_{-}u_{n}=A_{-}(n)u_{n}+B_{-}(n)u_{n-1}+C_{-}u_{n+1}\text{,}  \label{k-1}
\end{equation}
\begin{equation}
K_{0}u_{n}=A_{0}(n)u_{n}+C_{0}u_{n+1}\text{,}  \label{k01}
\end{equation}
where $C_{+}=\frac{1}{2}$, $C_{-}=2\rho ^{2}$, $C_{0}=\rho $, $%
A_{-}(n)=(2\beta +4n+1)\rho $, $A_{0}(n)=\frac{2\beta +4n+1}{4}$, $B_{-}(n)=%
\frac{(\beta +2n)(\beta +2n-1)}{2}$.

Using eqs.(\ref{c1}) - (\ref{k01}), one can present the action of the
operator (\ref{hk}) in the form

\begin{equation}
Hu_{n}=D_{2}u_{n+2}+D_{1}(n)u_{n+1}+\tilde{D}_{0}(n)u_{n}+\tilde{D}%
_{1}(n)u_{n-1}+\tilde{D}_{2}(n)u_{n-2}\text{,}
\end{equation}
where 
\begin{eqnarray}
D_{2} &=&\frac{a_{+0}}{2}%
C_{0}+a_{0-}C_{-}C_{0}+a_{00}C_{0}^{2}+a_{--}C_{-}C_{-}+\frac{a_{++}}{4}
\label{m1} \\
D_{1}\left( n\right) &=&a_{--}\left[ A_{-}(n)C_{-}+C_{-}A_{-}(n+1)\right]
+a_{00}\left[ A_{0}(n)C_{0}+C_{0}A_{0}(n+1)\right] +\frac{a_{+0}}{2}A_{0}(n)+
\\
&&a_{0-}[A_{-}(n)C_{0}+C_{-}A_{0}(n+1)]+a_{0}C_{0}+a_{-}C_{-}+\frac{a_{+}}{2}%
\text{,}  \nonumber \\
\tilde{D}_{0}
&=&a_{00}A_{0}^{2}(n)+a_{0-}[A_{-}(n)A_{0}(n)+B_{-}(n)C_{0}]+a_{0}A_{0}(n)+a_{-}A_{-}(n)%
\text{,} \\
\tilde{D}_{1}\left( n\right) &=&a_{--}\left[ A_{-}(n)B_{-}\left( n\right)
+B_{-}\left( n\right) A_{-}(n-1)\right]
+a_{0-}B_{-}(n)A_{0}(n-1)+a_{-}B_{-}(n)\text{,} \\
\tilde{D}_{2} &=&a_{--}B_{-}\left( n\right) B_{-}\left( n-1\right) \text{.}
\label{m6}
\end{eqnarray}
For the operator (\ref{hk}) to be quasi-exactly solvable with the invariant
subspace (\ref{f1}), it is necessary that the following conditions of cut
off be satisfied:

\begin{eqnarray}
D_{2} &=&0\text{,}  \label{cut1} \\
D_{1}(N) &=&0\text{,} \\
\tilde{D}_{1}(0) &=&0\text{,} \\
\tilde{D}_{2}(0) &=&0\text{,} \\
\tilde{D}_{2}(1) &=&0\text{.}  \label{cut5}
\end{eqnarray}
In general, this system is rather cumbersome. However, it is simplified
greatly if we consider Hermitian Hamiltonians with $a_{--}=0=a_{++}$. Then $%
\tilde{D}_{2}\equiv 0$ and we get three equations 
\begin{equation}
\rho ^{2}a_{00}+2\rho ^{3}a_{0-}+\frac{1}{2}\rho a_{0-}=0
\end{equation}
\begin{equation}
\lbrack \frac{a_{0-}}{4}(2\beta -3)+a_{-}]\beta (\beta -1)=0
\end{equation}
\begin{equation}
\frac{1+4\rho ^{2}}{2}a_{-}+\rho a_{0}+[\frac{2\beta +4N+1}{8}+\rho ^{2}%
\frac{7+6\beta +12N}{2}]a_{0-}+\frac{a_{00}}{2}\rho (2\beta +4N+3)=0.
\end{equation}
It is assumed that $a_{+}=a_{-}$, $a_{+0}=a_{0-}$, all coefficients are
real. The analysis leads to the following table of possible solutions:

$a_{++}=0$

\begin{tabular}{|l|l|l|l|l|l|}
\hline
& $\rho $ & $\beta $ & $a_{-}$ & $a_{00}$ & $a_{0}$ \\ \hline
1 & $0$ & $0$ & $-\frac{4N+1}{4}a_{0-}$ & a.v. & a.v. \\ \hline
2 & $0$ & $1$ & $-\frac{4N+3}{4}a_{0-}$ & a.v. & a.v. \\ \hline
3 & a.v. & a.v. & $\frac{3-2\beta }{4}a_{0-}$ & $f(\rho )a_{0-}$ & $[\frac{%
2\beta +2N+1}{4\rho }-2(N+1)\rho ]a_{0-}$ \\ \hline
\end{tabular}

$a_{++}\neq 0$

\begin{tabular}{|l|l|l|l|l|l|}
\hline
4 & a.v. & $0$ & a.v. & $f(\rho )a_{0-}$ & $f(\rho )a_{-}+f_{1}(\rho
,N)a_{0-}+f_{2}(\rho ,N)a_{++}$ \\ \hline
5 & a.v. & $1$ & a.v. & $f(\rho )a_{0-}$ & $f(\rho )a_{-}+f_{1}(\rho ,N+%
\frac{1}{2})a_{0-}+f_{2}(\rho ,N+\frac{1}{2})a_{++}$ \\ \hline
\end{tabular}

Here ''a.v.'' denotes ''arbitrary value'' with the reservation that $\beta $
is a positive integer or zero and $\left| \rho \right| <1/2$, as is
explained above. By definition,

\begin{equation}
f\left( \rho \right) =-\frac{1+4\rho ^{2}}{2\rho }\text{, }f_{1}(\rho ,N)=%
\frac{4N+5}{8\rho }-\frac{4N+1}{2}\rho \text{, }f_{2}(\rho ,N)=-\frac{4N+3}{%
8\rho ^{2}}(16\rho ^{4}-1)\text{.}  \label{f}
\end{equation}

One can check that the solution 1, when $u=1$, corresponds to even states
for the example considered in eq. (7) of \cite{bose} provided in that
equation the coefficient $A_{2}=0$. In a similar way, the case 2 ($u=z$)
corresponds to odd states from the same example. However, the cases (3)-(5)
represent new solutions that were not contained in \cite{bose}.

\section{doubled invariant subspaces}

Much more rich family of new classes of Bose QES quartic Hamiltonians can be
obtained if we generalize the structure of the invariant subspace
introducing, in addition to $u_{n}$, a second subset of independent
functions. \bigskip Consider the set of functions 
\begin{equation}
u_{n}=z^{2n}u\text{, }v_{n}=z^{2n+1}v\text{, }n=0,1,2...  \label{set}
\end{equation}
We are interested in such function which form a set, defined for a fixed $%
N=0,1,2$..., 
\begin{eqnarray}
F &=&span\left\{ u_{n},v_{n}\right\} \text{{}}=span\left\{ z^{2n}\cdot
u\left( z\right) ,z^{2n+1}\cdot v\left( z\right) \right\} \text{{}, }
\label{FN} \\
n &=&0\text{,}1\text{,}2\text{,...,}N\text{ ;}  \nonumber
\end{eqnarray}
invariant with respect to the action of the operator $H$. The dimension of $%
F $ is equal to $2(N+1)$. It may happen that, for some values of parameters,
the functions $v_{n}$ may be proportional or even exactly equal to $u_{n}.$
Then our subspace reduces to the $N+1$ one (\ref{f1}) considered in a
previous section. To gain qualitatively new QES models, in what follows we
will consider the functions $u_{n}$ and $v_{n}$ as, generally speaking,
independent.

Let $u$ and $v\,$obey the system of differential equations that generalizes
the relation (\ref{ud}): 
\begin{eqnarray}
u^{\prime } &=&Au+Bv\text{,}  \label{u} \\
v^{\prime } &=&Cu+Dv\text{.}  \label{v}
\end{eqnarray}
Here prime denotes differentiation with respect to $z$, $A$, $B$, $C$, $D$
are functions of $z$.

It follows from (\ref{u}), (\ref{v}) that 
\begin{equation}
u^{\prime \prime }-u^{\prime }\left( S+\frac{B^{\prime }}{B}\right)
+u[\Delta +\frac{W(B,A)}{B}]=0\text{,}  \label{u''}
\end{equation}
\begin{equation}
v^{\prime \prime }-v^{\prime }\left( S+\frac{C^{\prime }}{C}\right)
+v[\Delta +\frac{W(C,D)}{C}]=0\text{.}  \label{v''}
\end{equation}
Here $S=A+D=SpL$, where $L=\left( 
\begin{array}{ll}
A & B \\ 
C & D
\end{array}
\right) $, $W(f_{1},f_{2})\equiv f_{1}^{\prime }f_{2}-f_{1}f_{2}^{^{\prime
}} $ is a Wronskian, $\Delta =AD-BC$ is a determinant of $L$. In what
follows we assume for simplicity that quantities in denominators in (\ref
{u''}), (\ref{v''}) $B(z)\equiv \alpha =const$ and $C(z)=\gamma =const$.
Then we have 
\begin{eqnarray}
u^{\prime \prime }-u^{\prime }S+u(\Delta -A^{\prime }) &=&0\text{,}
\label{u1} \\
v^{\prime \prime }-v^{\prime }S+v(\Delta -D^{\prime }) &=&0\text{.}
\label{v1}
\end{eqnarray}
We assume also, by analogy with (\ref{ud}), that $A$ and $D$ contain only
terms of the order $z$ and $z^{-1}$ in the Loran series: $A=2\rho z+\beta
z^{-1}$, $D=2\tau z+\delta z^{-1}$. Then we have 
\begin{eqnarray}
\frac{d}{dz}u\left( z\right) &=&\alpha v\left( z\right) +\frac{\beta }{z}%
u\left( z\right) +2\rho zu\left( z\right) \text{,}  \label{sys} \\
\frac{d}{dz}v\left( z\right) &=&\gamma u\left( z\right) +\frac{\delta }{z}%
v\left( z\right) +2\tau zv\left( z\right) \text{.}  \nonumber
\end{eqnarray}

One obtains from (\ref{u1}), (\ref{v1}) 
\begin{equation}
u^{\prime \prime }-u^{\prime }[\frac{\delta +\beta }{z}+2(\rho +\tau )z]+u[%
\frac{\beta (1+\delta )}{z^{2}}+4\rho \tau z^{2}+2\beta \tau +2\rho \delta
-2\rho -\alpha \gamma ]=0\text{,}  \label{uu}
\end{equation}

\begin{equation}
v^{\prime \prime }-v^{\prime }[\frac{\delta +\beta }{z}+2(\rho +\tau )z]+v[%
\frac{\delta (1+\beta )}{z^{2}}+4\rho \tau z^{2}+2\tau \beta +2\rho \delta
-2\tau -\alpha \gamma ]=0\text{.}  \label{vv}
\end{equation}
Our functions $u_{n}(z)$ and $v_{n}(z)$ must belong to the Bargmann-Fock
space that entails, similarly to what is obtained in the previous section,
the conditions $\left| \rho \right| <1/2$, $\left| \tau \right| <1/2$.%
{\large \ }

To elucidate what constraints are imposed by the demand of analyticity,
consider separately several different cases. If $\gamma =\alpha =0$, eqs. (%
\ref{sys}) can be integrated and one easily finds that $u=z^{\beta }\exp
(\rho z^{2})$, $v=z^{\delta }\exp (\tau z^{2})$, whence it is obvious that $%
\beta =0,1,2...$ and $\delta =-1,0,1,2...$ If $\alpha =0$ but $\gamma \neq 0$%
, one can make the substitution $v=z^{\delta }\exp (\tau z^{2})w$. Then 
\begin{equation}
w^{\prime }=\gamma z^{\beta -\delta }\exp [(\rho -\tau )z^{2}]  \label{w}
\end{equation}
It is clear that $\beta =0$, $1$, $2$..., whereas $\delta $ is arbitrary
except $\delta =\beta +1$ since the latter would have led to the logarithmic
terms in $v(z)$. The similar situation occurs when $\gamma =0$ but $\alpha
\neq 0$. Then $\delta =-1,0$, $1$, $2$... and forbidden values of $\beta $
are $\beta =\delta +1$.

Let now $\alpha \gamma \neq 0$. First, consider the case $\rho \neq \tau $.
Then by substitutions 
\begin{equation}
u(z)=y\left( z^{2}\left( \tau -\rho \right) \right) \cdot \exp \left( \rho
z^{2}\right) \cdot z^{\beta }  \label{up}
\end{equation}
\begin{equation}
v(z)=\tilde{y}\left( z^{2}\left( \rho -\tau \right) \right) \cdot \exp
\left( \tau z^{2}\right) \cdot z^{\delta }  \label{vp}
\end{equation}
eqs. (\ref{uu}), (\ref{vv}) are reduced to the form, typical of a degenerate
hypergeometric function 
\begin{equation}
x\frac{d^{2}}{dx^{2}}y\left( x\right) +\left( \eta -x\right) \frac{d}{dx}%
y\left( x\right) -\xi y\left( x\right) =0\text{,}  \label{eq1}
\end{equation}
where $\eta =\frac{1}{2}\left( \beta -\delta +1\right) ,\xi =\frac{\alpha
\gamma }{4\left( \tau -\rho \right) }$. The function $\tilde{y}$ satisfies
the equation of the same form (\ref{eq1}) but with parameters $\tilde{\eta}%
=1-\eta $, $\tilde{\xi}=-\xi $.

To determine the admissible range of parameters $\beta $, $\delta $ one can
appeal directly to the well-known properties of this function and take into
account that the general solution of eq. (\ref{eq1}) has the form $%
y=Ay_{1}+By_{2}$, where $y_{1}=\Phi (\xi ,\eta ;x)$ and $y_{2}=$ $x^{1-\eta
}\Phi (\xi -\eta +1,2-\eta ;x)$ and the standard notation for the degenerate
hypergeometric function is used (see Ch. 6 of Ref. \cite{ba}). First
consider the case when $\eta $ is non-integer. Then $\Phi \rightarrow 1$
when $x\rightarrow 0$ and from (\ref{up}) we obtain the function $u$ can
have two possible asymptotic forms: $u_{1}\sim z^{\beta }$ and $u_{2}\sim
z^{\delta +1}$. The function $v(z)$ behaves, correspondingly, like $%
v_{1}\sim z^{\beta +1}$ and $v_{2}\sim z^{\delta }$. Therefore, it turns out
that there are two cases: 
\begin{equation}
\beta =0,1,2...,\delta \text{ is arbitrary}  \label{case1}
\end{equation}
or 
\begin{equation}
\delta =-1,0,2...,\beta \text{ is arbitrary}.  \label{case2}
\end{equation}

If $\eta $ is integer, there exists only one independent solution of eq. (%
\ref{eq1}), regular at $x\rightarrow 0$. The corresponding solution is known
to be $\Phi ^{*}(\xi ,\eta ;x)=\frac{\Phi (\xi ,\eta ;x)}{\Gamma (\eta )}$ .
In the limit $x\rightarrow 0$ $\Phi ^{*}\sim x^{1-\eta }$ that does not
affect the conclusion about admissible range of $\beta $ and $\delta $.

Let now $\rho =\tau $, $\alpha \gamma \neq 0$. By substitution 
\[
u(z)=y\left( 2\sqrt{\alpha \gamma }z\right) \cdot \exp \left( z\left( \rho z-%
\sqrt{\alpha \gamma }\right) \right) \cdot z^{\beta } 
\]
we obtain that the function $y(x)$ obeys the equation 
\[
x\frac{d^{2}}{dx^{2}}y\left( x\right) +\left( \beta -\delta -x\right) \frac{d%
}{dx}y\left( x\right) -\frac{(\beta -\delta )}{2}y\left( x\right) =0 
\]
that has the same form as (\ref{eq1}) and admissible $\beta $ and $\delta $
satisfy one of criteria (\ref{case1}), (\ref{case2}).

Differential equations for our functions can be also written in the
symmetric form. Let us make the substitution 
\begin{equation}
u=\Psi z^{\frac{\beta +\delta }{2}}\exp [\frac{(\rho +\tau )}{2}z^{2}]\text{.%
}  \label{sub}
\end{equation}
Then 
\begin{equation}
\Psi ^{\prime \prime }+(\varepsilon -V_{eff})\Psi =0\text{,}  \label{sc}
\end{equation}
where {\large \ } 
\begin{equation}
V_{eff}=\frac{k}{z^{2}}+(\rho -\tau )^{2}z^{2}\text{,}  \label{ef}
\end{equation}
\begin{equation}
k=\frac{(\delta -\beta )(\delta -\beta +2)}{4}\text{.}  \label{l}
\end{equation}
\begin{equation}
\varepsilon =(\rho -\tau )(\delta -\beta -1)-\alpha \gamma .  \label{e}
\end{equation}
In a similar way, 
\begin{equation}
v=\tilde{\Psi}z^{\frac{\beta +\delta }{2}}\exp [\frac{(\rho +\tau )}{2}z^{2}]%
\text{,}  \label{vs}
\end{equation}
where $\tilde{\Psi}$ obeys the equation (\ref{sc}) with the same structure
of $V_{eff}$ but with another $\tilde{k}=\frac{(\beta -\delta )(\beta
-\delta +2)}{4}$. $\tilde{\varepsilon}=(\tau -\rho )(\beta -\delta
-1)-\alpha \gamma .$ It is seen that $\tilde{\varepsilon}$ can be obtained
from $\varepsilon $ by interchange between $\rho $ and $\tau $, $\beta $ and 
$\delta $. It is also seen that $\tilde{\varepsilon}-\varepsilon =2(\rho
-\tau )$.

Thus, formally, we obtain the harmonic oscillator with a barrier $z^{-2}$
(Kratzer Hamiltonian). Let us remind, however, that the variable $z$ in our
context is complex.

It follows from (\ref{sys}) and (\ref{sub}), (\ref{vs}) that the functions $%
\Psi $, $\tilde{\Psi}$ obey the system of equations 
\begin{equation}
\left[ \frac{d}{dz}+(\tau -\rho )z+\frac{\delta -\beta }{2z}\right] \Psi
=\alpha \tilde{\Psi}\text{,}  \label{a}
\end{equation}
\begin{equation}
\left[ \frac{d}{dz}+(\rho -\tau )z+\frac{\beta -\delta }{2z}\right] \tilde{%
\Psi}=\gamma \Psi \text{.}  \label{a+}
\end{equation}

\section{conditions of quasi-exact solvability}

{\large \bigskip }The action of operators $K_{i}$ in the subspace (\ref{FN})
has the same structure (\ref{c1}) - (\ref{k01}) but now the corresponding
quantities $A_{i}$, $B_{i}$, $C_{i}$ becomes $2\times 2$ matrices:

\begin{equation}
K_{+}\vec{f}_{n}=C_{+}\vec{f}_{n+1}\text{,}  \label{k+}
\end{equation}
\begin{equation}
K_{-}\vec{f}_{n}=A_{-}(n)\vec{f}_{n}+B_{-}(n)\vec{f}_{n-1}+C_{-}\vec{f}_{n+1}%
\text{,}  \label{k-}
\end{equation}
\begin{equation}
K_{0}\vec{f}_{n}=A_{0}(n)\vec{f}_{n}+C_{0}\vec{f}_{n+1}\text{,}  \label{k0}
\end{equation}
where $\vec{f}_{n}=\left( 
\begin{array}{l}
u_{n} \\ 
v_{n}
\end{array}
\right) $, 
\begin{equation}
C_{+}=\frac{1}{2}\left( 
\begin{array}{ll}
1 & 0 \\ 
0 & 1
\end{array}
\right) \text{, }C_{-}=\left( 
\begin{array}{ll}
2\rho ^{2} & 0 \\ 
\gamma (\tau +\rho ) & 2\tau ^{2}
\end{array}
\right) \text{, }C_{0}=\frac{1}{2}\left( 
\begin{array}{ll}
2\rho & 0 \\ 
\gamma & 2\tau
\end{array}
\right) \text{,}  \label{c}
\end{equation}
\begin{equation}
A_{-}(n)=\frac{1}{2}\left( 
\begin{array}{ll}
4\beta \rho +8n\rho +2\rho +\alpha \gamma & 2\alpha (\tau +\rho ) \\ 
\gamma (\beta +\delta +2+4n) & 4\delta \tau +8n\tau +6\tau +\alpha \gamma
\end{array}
\right) \text{,}  \label{a-}
\end{equation}
\begin{equation}
A_{0}(n)=\frac{1}{4}\left( 
\begin{array}{ll}
2\beta +4n+1 & 2\alpha \\ 
0 & 2\delta +4n+3
\end{array}
\right) \text{,}  \label{a0}
\end{equation}
\begin{equation}
B_{-}(n)=\frac{1}{2}\left( 
\begin{array}{ll}
(\beta +2n)(\beta +2n-1) & \alpha (\beta +\delta +4n) \\ 
0 & (\delta +2n+1)(\delta +2n)
\end{array}
\right) \text{.}  \label{b-}
\end{equation}

In a similar way, the action of Hamiltonian in the invariant subspace can be
represented in the form

\begin{equation}
H\vec{f}_{n}=D_{2}\vec{f}_{n+2}+D_{1}(n)\vec{f}_{n+1}+\tilde{D}_{0}(n)\vec{f}%
_{n}+\tilde{D}_{1}(n)\vec{f}_{n-1}+\tilde{D}_{2}(n)\vec{f}_{n-2}\text{,}
\label{hd}
\end{equation}
where matrices $D_{i}$ and $\tilde{D}_{i}$ have the form (\ref{m1}) - (\ref
{m6}) with $A_{i}$, $B_{i}$, $C_{i}$ taken from eqs. (\ref{c}) - (\ref{b-})
(the order of operators is taken into account properly in this form of
writing).

For the operator (\ref{hk}) to be quasi-exactly solvable with the invariant
subspace (\ref{FN}), it is necessary that the matrix version of the
conditions of cut off (\ref{cut1}) - (\ref{cut5}) be satisfied. The
corresponding system of equations is too cumbersome to be listed here. It
can be simplified greatly if we assume the condition $a_{--}=0$, in which
case after simple calculations we get $\tilde{D}_{2}=0$, 
\begin{equation}
D_{2}=\left( 
\begin{array}{ll}
\rho ^{2}a_{00}+2\rho ^{3}a_{0-}+\frac{1}{2}\rho a_{+0}+\frac{a_{++}}{4} & 0
\\ 
\gamma Y & \tau ^{2}a_{00}+2\tau ^{3}a_{0-}+\frac{1}{2}\tau a_{+0}+\frac{%
a_{++}}{4}
\end{array}
\right)  \label{D2m}
\end{equation}

where $Y\equiv [\frac{1}{2}(\rho +\tau )a_{00}+(\rho ^{2}+\tau ^{2}+\rho
\tau )a_{0-}+\frac{1}{4}a_{+0}]$, 
\begin{equation}
2\tilde{D}_{1}(0)=\left( 
\begin{array}{ll}
\lbrack \frac{a_{0-}}{4}(2\beta -3)+a_{-}]\beta (\beta -1) & \alpha [\frac{%
a_{0-}}{4}(2\beta ^{2}+2\delta ^{2}+2\beta \delta -3\beta -\delta
)+a_{-}(\beta +\delta )] \\ 
0 & \frac{\delta (\delta +1)}{2}[(2\delta -1)\frac{a_{0-}}{4}+a_{-}]
\end{array}
\right)  \label{D-}
\end{equation}
\begin{equation}
D_{1}(N)=\left( 
\begin{array}{ll}
d_{1} & d_{2} \\ 
d_{3} & d_{4}
\end{array}
\right)  \label{dn}
\end{equation}
\begin{eqnarray}
d_{1} &=&\frac{a_{+}}{2}+2\rho ^{2}a_{-}+\rho a_{0}+\frac{a_{+0}}{8}(2\beta
+4N+1)+\frac{a_{0-}}{2}[\alpha \gamma (\tau +2\rho )+\rho ^{2}(7+6\beta
+12N)]  \label{d1} \\
&&+\frac{a_{00}}{4}[\rho (4\beta +8N+6)+\alpha \gamma ]\text{,}
\end{eqnarray}
\begin{equation}
d_{2}=\alpha Y\text{,}  \label{d2y}
\end{equation}
\begin{equation}
d_{3}=\frac{\gamma }{2}Z\text{, }Z=[2(\tau +\rho )a_{-}+a_{0}+\frac{a_{0-}}{2%
}\xi +\frac{a_{00}}{2}(\beta +\delta +4N+4)]\text{,}  \label{d3}
\end{equation}
\begin{equation}
\xi =12N(\rho +\tau )+\alpha \gamma +\tau (4\delta +2\beta +11)+\rho (4\beta
+2\delta +9)\text{,}  \label{x}
\end{equation}
\begin{eqnarray}
d_{4} &=&\frac{a_{+}}{2}+2\tau ^{2}a_{-}+\tau a_{0}+\frac{a_{+0}}{8}(2\delta
+4N+3)+\frac{a_{0-}}{2}[\alpha \gamma (\rho +2\tau )+\tau
^{2}(12N+13+6\delta )]  \label{d4} \\
&&+\frac{a_{00}}{4}[\tau (4\delta +8N+10)+\alpha \gamma ]
\end{eqnarray}

One can observe that 
\begin{equation}
d_{1}-d_{4}=(\rho -\tau )Z+(\beta -\delta -1)Y  \label{obs}
\end{equation}
The system of equations (\ref{cut1}) - (\ref{cut5}) with (\ref{D2m}) - (\ref
{d4}) taken into account looks rather cumbersome but the relation (\ref{obs}%
) simplifies analysis significantly.

However, for a generic case $a_{--}\neq 0$ algebraic calculation are so
bulky that we had to resort to using a computer.

In what follows we restrict ourselves by the Hermitian case only which is
the most interesting for physical applications. This implies that $%
a_{++}=a_{--}$, $a_{+}=a_{-},a_{+0}=a_{0-}$, where all coefficients are real.

The full set of non-trivial Hermitian solutions of the system (\ref{cut1}) -
(\ref{cut5}){\large \ }and their classification are given in the next
Section.

\section{ Hermitian solutions of algebraic equations}

For the type of invariant subspaces (\ref{FN}) under consideration, we
suggest below the classification of {\it all} QES Hermitian Hamiltonians,
quadratic-linear with respect to operators $K_{i}$ (i.e., those which
represent even polynomial of the fourth order in terms of $a$, $a^{+}$).

In the tables below we list only qualitatively different cases in the
following sense. If some solutions can be obtained by the limiting
transition ($a_{++}\rightarrow 0$, $\gamma \rightarrow 0$, etc.) from a more
general case, we do not repeat them. As before, we use abbreviation ''a.v.''
for ''arbitrary value''.

\subsection{a$_{++}=0$}

$\gamma \neq 0$

\begin{tabular}{|c|c|c|c|c|c|c|c|c|}
\hline
& $\delta $ & $\beta $ & $\alpha $ & $\rho $ & $\tau $ & $a_{-}$ & $a_{00}$
& $a_{0}$ \\ \hline
1 & a.v. & $0$ & a.v. & $0$ & $\frac{\alpha \gamma }{2(2N+1-\delta )}$ & $%
\frac{1-2\delta }{4}a_{0-}$ & $f\left( \tau \right) a_{0-}$ & $[\tau g\left(
\delta \right) +\frac{\delta +4N+4}{4\tau }]a_{0-}$ \\ \hline
2 & a.v. & $0$ & a.v. & $\frac{\alpha \gamma }{4(N+1)}$ & $0$ & $\frac{%
1-2\delta }{4}a_{0-}$ & $f\left( \rho \right) a_{0-}$ & $[\rho g\left(
\delta \right) +\frac{\delta +4N+4}{4\rho }]a_{0-}$ \\ \hline
3 & a.v. & $1$ & a.v. & $0$ & $\frac{\alpha \gamma }{2(2N+2-\delta )}$ & $%
\frac{1-2\delta }{4}a_{0-}$ & $f\left( \tau \right) a_{0-}$ & $[\tau g\left(
\delta -1\right) +\frac{\delta +4N+5}{4\tau }]a_{0-}$ \\ \hline
4 & a.v. & $1$ & a.v. & $\frac{\alpha \gamma }{4(N+1)}$ & $0$ & $\frac{%
1-2\delta }{4}a_{0-}$ & $f\left( \rho \right) a_{0-}$ & $[\rho g\left(
\delta -1\right) +\frac{\delta +4N+5}{4\rho }]a_{0-}$ \\ \hline
5 & $0$ & a.v. & a.v. & $0$ & $\frac{\alpha \gamma }{4(N+1)}$ & $\frac{%
3-2\beta }{4}a_{0-}$ & $f\left( \tau \right) a_{0-}$ & $[\tau g\left( \beta
-2\right) +\frac{\beta +4N+4}{4\tau }]a_{0-}$ \\ \hline
6 & $0$ & a.v. & a.v. & $\frac{\alpha \gamma }{2(2N+3-\beta )}$ & $0$ & $%
\frac{3-2\beta }{4}a_{0-}$ & $f\left( \rho \right) a_{0-}$ & $[\rho g\left(
\beta -2\right) +\frac{\beta +4N+4}{4\rho }]a_{0-}$ \\ \hline
7 & $0$ & $0$ & a.v. & $0$ & a.v. & $\frac{\alpha \gamma -\tau (4N+1)}{4\tau 
}a_{0-}$ & $f\left( \tau \right) a_{0-}$ & $(-\tau -\alpha \gamma +\frac{N+1%
}{\tau })a_{0-}$ \\ \hline
8 & $0$ & $0$ & a.v. & a. v. & $0$ & $\frac{\alpha \gamma -\rho (4N+3)}{%
4\rho }a_{0-}$ & $f\left( \rho \right) a_{0-}$ & $(\rho -\alpha \gamma +%
\frac{N+1}{\rho })a_{0-}$ \\ \hline
9$^{*}$ & $0$ & $0$ & a.v. & $0$ & $0$ & $-\frac{\alpha \gamma }{2}a_{00}$ & 
a. v. & $-2(N+1)a_{00}$ \\ \hline
10 & $0$ & $1$ & $0$ & $0$ & a. v. & $-\frac{4N+3}{4}a_{0-}$ & $f\left( \tau
\right) a_{0-}$ & $\frac{4N+5}{4\tau }a_{0-}$ \\ \hline
11 & $0$ & $1$ & $0$ & a.v. & $0$ & $-\frac{4N+3}{4}a_{0-}$ & $f\left( \rho
\right) a_{0-}$ & $\frac{4N+5}{4\rho }a_{0-}$ \\ \hline
12 & $-1$ & $1$ & a.v. & $0$ & a.v. & $\frac{\alpha \gamma -\tau (4N+3)}{%
4\tau }a_{0-}$ & $f\left( \tau \right) a_{0-}$ & $\left( \tau -\alpha \gamma
+\frac{N+1}{\tau }\right) a_{0-}$ \\ \hline
13 & $-1$ & $1$ & a.v. & a.v. & $0$ & $\frac{\alpha \gamma -\rho (4N+1)}{%
4\rho }a_{0-}$ & $f\left( \rho \right) a_{0-}$ & $(-\rho -\alpha \gamma +%
\frac{N+1}{\rho })a_{0-}$ \\ \hline
14 & $-1$ & $1$ & a.v. & $0$ & $0$ & $-\frac{\alpha \gamma }{2}a_{00}$ & a.
v. & $-2(N+1)a_{00}$ \\ \hline
15 & $-1$ & $0$ & $0$ & $0$ & a.v. & $-\frac{4N+1}{4}a_{0-}$ & $f\left( \tau
\right) a_{0-}$ & $\frac{4N+3}{4\tau }a_{0-}$ \\ \hline
16 & $-1$ & $0$ & $0$ & a.v. & $0$ & $-\frac{4N+1}{4}a_{0-}$ & $f\left( \rho
\right) a_{0-}$ & $\frac{4N+3}{4\rho }a_{0-}$ \\ \hline
17 & $-1$ & a.v. & a.v. & $0$ & $\frac{\alpha \gamma }{4(N+1)}$ & $\frac{%
3-2\beta }{4}a_{0-}$ & $f\left( \tau \right) a_{0-}$ & $[\tau g\left( \beta
-1\right) +\frac{\beta +4N+3}{4\tau }]a_{0-}$ \\ \hline
18 & $-1$ & a.v. & a.v. & $\frac{\alpha \gamma }{2(2N+2-\beta )}$ & $0$ & $%
\frac{3-2\beta }{4}a_{0-}$ & $f\left( \rho \right) a_{0-}$ & $[\rho g\left(
\beta -1\right) +\frac{\beta +4N+3}{4\rho }]a_{0-}$ \\ \hline
\end{tabular}

where 
\begin{equation}
g\left( x\right) \doteqdot x-3-4N  \label{g}
\end{equation}
and we used definition of $f$ according to (\ref{f}).

$^{*}$Note: in the case 9 the coefficients $a_{+0}=a_{0-}=0$.

$\gamma =0$

\begin{tabular}{|l|l|l|l|l|l|l|l|l|}
\hline
& $\delta $ & $\beta $ & $\alpha $ & $\rho $ & $\tau $ & $a_{-}$ & $a_{00}$
& $a_{0}$ \\ \hline
19 & a.v. & $\delta +1$ & $0$ & a.v. & $\rho $ & $\frac{1-2\delta }{4}a_{0-}$
& $f\left( \rho \right) a_{0-}$ & $a_{0-}[\frac{2\delta +2N+3}{4\rho }-2\rho
(N+1)]$ \\ \hline
20 & $0$ & $1$ & $0$ & $0$ & $0$ & $-\frac{(4N+3)}{4}a_{0-}$ & a.v. & a.v.
\\ \hline
21 & $-1$ & $0$ & $0$ & $0$ & $0$ & $-\frac{(1+4N)}{4}a_{0-}$ & $0$ & $0$ \\ 
\hline
\end{tabular}

\subsection{a$_{++}\neq 0$}

$\gamma \neq 0$

In all admissible cases 22 - 25 $\rho $ and $\tau $ take arbitrary values, 
\[
a_{0-}=-\frac{1}{2\rho \,\tau }\left( \tau +\rho \right) \left( 4\,{\tau }%
\rho +1\right) a_{+\text{ }+},a_{00}=\frac{1}{4\rho \tau }\left( \left( 4\,{%
\tau }\rho +1\right) ^{2}+4\left( \,{\rho }^{2}+\,{\tau }^{2}\right) \right)
a_{+\text{ }+}\text{.} 
\]

The rest of relevant quantities is

\begin{tabular}{|l|l|l|l|l|l|}
\hline
$^{{}}$ & $\delta $ & $\beta $ & $\alpha $ & $a_{0}$ & $a_{-}$ \\ \hline
22 & $0$ & $0$ & a.v. & $f_{0}(\rho ,\tau )\,a_{+\text{ }+}$ & $f_{-}(\rho
,\tau )a_{++}$ \\ \hline
23 & $0$ & $1$ & $0$ & $g_{0}(\rho ,\tau ,N)a_{+\text{ }+}$ & $g_{-}(\rho
,\tau ,N)a_{+\text{ }+}$ \\ \hline
24 & $-1$ & $1$ & a.v. & $f_{0}(\rho ,\tau )\,a_{+\text{ }+}$ & $f_{-}(\tau
,\rho )a_{++}$ \\ \hline
25 & $-1$ & $0$ & $0$ & $g_{0}(\rho ,\tau ,N-\frac{1}{2})a_{+\text{ }+}$ & $%
g_{-}(\rho ,\tau ,N-\frac{1}{2})a_{+\text{ }+}$ \\ \hline
\end{tabular}

Here $f_{0}(\rho ,\tau )\equiv \frac{1}{2\rho \,\tau }[\left( 16{\tau }^{2}{%
\rho }^{2}-1\right) \left( N+1\right) -{\rho }^{2}+{\tau }^{2}+\alpha \gamma
\left( \tau +\rho \right) ]$,

$g_{0}(\rho ,\tau ,N)=\frac{\left( 5+4\,N\right) }{8\rho \tau }\left( 16{%
\tau }^{2}{\rho }^{2}-1\right) $, $g_{-}(\rho ,\tau ,N)=-\frac{\left( \tau
+\rho \right) }{8\rho \tau }\left[ {\tau }\rho (16N+28)-4N-3\right] $

$f_{-}(\rho ,\tau )\equiv -\frac{1}{8\rho \,\tau }[{\tau }^{2}\rho
(16N+28)+\tau \,{\rho }^{2}(20+16N)+\alpha \gamma (4\rho \tau +1)-\rho
(4N+3)-\tau (1+4\,N)]$

$\gamma =0$

Now $\rho =\tau $.

\begin{tabular}{|l|l|l|l|l|l|l|}
\hline
& $\delta $ & $\beta $ & $\alpha $ & $a_{00}$ & $a_{0-}$ & $a_{0}$ \\ \hline
26 & $0$ & $0$ & a.v. & $f_{00}(\tau )a_{+\text{ }+}$ & $2f(\tau )a_{+\text{ 
}+}$ & $-f^{2}(\tau )a_{++}+f(\tau )a_{-}$ \\ \hline
27 & $0$ & $1$ & $0$ & $f(\tau )a_{0-}+f(2\tau ^{2})a_{++}$ & a.v. & $f(\tau
)a_{-}-g_{0}(\tau ,\tau ,N)a_{++}+f_{1}(\tau ,N+\frac{1}{2})a_{0-}$ \\ \hline
28 & $-1$ & $1$ & a.v. & $f_{00}(\tau )a_{+\text{ }+}$ & $2f(\tau )a_{+\text{
}+}$ & $-f^{2}(\tau )a_{++}+f(\tau )a_{-}$ \\ \hline
29 & $-1$ & $0$ & $0$ & $f(\tau )a_{0-}+f(2\tau ^{2})a_{++}$ & a.v. & $%
f(\tau )a_{-}-g_{0}(\tau ,\tau ,N-\frac{1}{2})a_{++}+f_{1}(\tau ,N)a_{0-}$
\\ \hline
\end{tabular}

\[
f_{00}(\tau )=\frac{1}{4\tau ^{2}}\left( 1+16{\tau }^{4}+16\,{\tau }%
^{2}\right) \text{,} 
\]
the function $f_{1}$ is defined according to (\ref{f}).

\section{explicit examples of invariant subspaces}

In this section we list shortly the explicit form of solutions for $u(z)$
and $v(z)$.

1) The case 9: $\beta =\delta =\rho =\tau =0$,

a) $\gamma =-\alpha =-\omega \neq 0$

Then it follows directly from (\ref{sys}) that the functions $u$, $v$ can be
chosen as $u=\cos \omega z$, $v=\sin \omega z$. After simple calculations
one finds that, apart from the Hermitian QES Hamiltonian, there exists also
the non-Hermitian QES operator: 
\begin{equation}
H=a_{00}K_{0}^{2}+a_{0-}K_{0}K_{-}+[\frac{a_{0-}}{2}\omega
^{2}-2a_{00}(N+1)]K_{0}+a_{-}K_{-}+\frac{a_{00}}{2}\omega ^{2}K_{+}
\label{h1}
\end{equation}

b) In a similar way, we obtain for $\gamma =\alpha =\omega $ that $u=\cosh
\left( \omega z\right) $, $v=\sinh \omega z$, and the operator $H$ is
obtained by replacement $\omega ^{2}\rightarrow -\omega ^{2}$ in the
expression (\ref{h1}).

2) $\rho =\tau =0$, $\delta =-1-n$, $\beta =n$ $(n=0$,$1...)$, $\alpha =-1$, 
$\gamma =1$

Then we have the following solutions of (\ref{sys}): $u=J_{n}(z)$, $%
v(z)=J_{n+1}(z)$ (Bessel functions).

\begin{equation}
a_{-}=\frac{(3+2n)}{4}a_{0-}\text{, }a_{+0}=0=a_{++}\text{, }a_{+}=\frac{%
a_{00}}{2}\text{, }a_{0}=\frac{a_{0-}}{2}-\frac{a_{00}}{2}(4N+3).
\label{bes}
\end{equation}
The effective Hamiltonian is non-Hermitian.

3) Consider the case 8 of Hermitian Hamiltonians (the case 7 can be
considered in a similar manner): $\beta =\delta =0=\tau $, then $k=0=\tilde{k%
}$ and formally we have in the $z$-representation the wave function that
looks like that of a pure harmonic oscillator would look in the coordinate
representation. Eqs. (\ref{a}), (\ref{a+}) take the form 
\begin{equation}
b\Psi =\frac{\alpha }{\sqrt{2\rho }}\tilde{\Psi}\text{,}  \label{a1}
\end{equation}
\begin{equation}
b^{+}\tilde{\Psi}=\frac{\gamma }{\sqrt{2\rho }}\Psi \text{,}  \label{a2}
\end{equation}
where $b=\frac{1}{\sqrt{2\rho }}[\frac{d}{dz}-\rho z]$, $b^{+}=\frac{1}{%
\sqrt{2\rho }}[\frac{d}{dz}+\rho z]$. It is obvious that $[b,b^{+}]=1$.

The frequency is equal to $\omega =2\rho $, $\varepsilon =-\frac{\omega }{2}%
-\alpha \gamma $, $\tilde{\varepsilon}-\varepsilon =2\rho $. Let also $%
\varepsilon =\varepsilon _{n}\equiv \omega (n+1/2)$, $\alpha \gamma =-\omega
(n+1)$. Then we have the eigenvalue and $\tilde{\varepsilon}=\varepsilon
_{n+1}$. Thus, our subspace is $span\{\Psi _{n}z^{2n}\exp [\left( \frac{\rho 
}{2}\right) z^{2}]$, $\Psi _{n+1}z^{2n+1}\exp [\left( \frac{\rho }{2}\right)
z^{2}]\}$, where $\Psi _{n}$ is the wave function of the n-th level of the
harmonic oscillator, $\Psi _{n}=\exp (-\frac{1}{2}\rho z^{2})H_{n}(z\sqrt{%
\rho })$, $H_{n}$ is the Hermite polynomial. Here $\beta =0$, $1$, ...

We would like to stress that our system represents an anharmonic (not
harmonic!) Bose oscillator. The functions, which have the same form as those
of an harmonic oscillator, appear in this context in the coherent state
representation (not in the coordinate one, as would be the case for the
usual harmonic oscillator) and represent auxiliary quantities.

\section{generalizations}

In these section we describe shortly, on the basis of the suggested
approach, some possible ways of generation of new invariant subspaces,
suitable for constructing QES Bose Hamiltonians. As the method of
constructing is the same as was used above, we only dwell upon the structure
of subspaces.

1) Let us introduce quantities 
\begin{equation}
f_{n}^{1}=z^{2n}u^{2}\text{, }f_{n}^{2}=z^{2n+1}uv\text{, }%
f_{n}^{3}=z^{2n}v^{2}
\end{equation}
and consider the subspace $F_{N}=span\left\{ f_{n}^{1},\text{ }f_{n}^{2},%
\text{ }f_{n}^{3}\right\} ${}, $n=0$,$1$,...,$N$ ; $N=1$,$2$,.... Let $\vec{f%
}_{n}\equiv \left( 
\begin{array}{l}
f_{n}^{1} \\ 
f_{n}^{2} \\ 
f_{n}^{3}
\end{array}
\right) $. Then (\ref{k+})-(\ref{k0}) take place, where, however, now the
corresponding matrices have dimension $3\times 3$: 
\begin{equation}
A_{0}(n)=\left( 
\begin{array}{lll}
\xi +\frac{1}{4} & \alpha & 0 \\ 
0 & \frac{2\xi +2\eta +3}{4} & 0 \\ 
0 & \gamma & \eta +\frac{1}{4}
\end{array}
\right) \text{, }C_{0}=\left( 
\begin{array}{lll}
2\rho & 0 & 0 \\ 
\frac{\gamma }{2} & \omega & \frac{\alpha }{2} \\ 
0 & 0 & 2\tau
\end{array}
\right) \text{, }C_{+}=\frac{1}{2}I\text{,}
\end{equation}
$I$ is a unit matrix, 
\begin{equation}
A_{-}(n)=\left( 
\begin{array}{lll}
2\rho \left( 4\xi +1\right) +\alpha \gamma & 2\alpha \left( \omega +2\rho
\right) & \alpha ^{2} \\ 
\frac{\gamma }{2}\left( 2+3\xi +\eta \right) & \omega \left( 2\xi +2\eta
+3\right) +2\alpha \gamma & \frac{\alpha }{2}\left( 2+\xi +3\eta \right) \\ 
\gamma ^{2} & 2\gamma \left( \omega +2\tau \right) & 2\tau \left( 1+4\eta
\right) +\alpha \gamma
\end{array}
\right) \text{,}
\end{equation}
\begin{equation}
B_{-}(n)=\left( 
\begin{array}{lll}
\xi \left( 2\xi -1\right) & \alpha \left( 3\xi +\eta \right) & 0 \\ 
0 & \frac{1}{2}\left( \eta +\xi +1\right) \left( \eta +\xi \right) & 0 \\ 
0 & \gamma \left( 3\eta +\xi \right) & \eta \left( 2\eta -1\right)
\end{array}
\right) \text{ }
\end{equation}
\begin{equation}
C_{-}(n)=\left( 
\begin{array}{lll}
8\rho ^{2} & 0 & 0 \\ 
\gamma \left( \omega +2\rho \right) & 2\omega ^{2} & \alpha \left( 2\tau
+\omega \right) \\ 
0 & 0 & 8\tau ^{2}
\end{array}
\right)
\end{equation}
$\xi =\beta +n$, $\eta =n+\delta $, $\omega =\rho +\tau $.

2) Further generalization of consists in considering 
\begin{equation}
f_{n}^{1}=z^{2n}u\tilde{u}\text{, }f_{n}^{2}=z^{2n+1}\tilde{u}v\text{, }%
f_{n}^{3}=z^{2n+1}u\tilde{v}\text{, }z^{2n+1}\tilde{u}v\text{, }%
f_{n}^{4}=z^{2n}v\tilde{v}
\end{equation}
Here $\tilde{u},\tilde{v}$ refer to the functions obeying eqs. (\ref{sys})
with parameters $\tilde{\alpha},\tilde{\beta},\tilde{\gamma},\tilde{\delta},%
\tilde{\rho},\tilde{\tau}$. Then we can construct $span\left\{ f_{n}^{1},%
\text{ }f_{n}^{2},\text{ }f_{n}^{3},f_{n}^{4}\right\} {}$.

3) Consider $span\{z^{2n+\xi \left( k\right) }u^{M-k}v^{k}\}$, $\xi (k)=1$,
if $k$ is odd and $\xi =0$, if $k$ is even. Here $k=0,1,2...M$, $n=0,1...N$.

The functions $\{u^{i}\}$ that appear in the subspaces 1)-3) obey the system
of equations of the type $\frac{du^{i}}{dz}=C_{j}^{i}\left( z\right) u^{j}$%
{\large \ }that contains, as a particular case, eqs. (\ref{sys}).

\section{Summary}

Let us summarize the basic features of our approach to constructing
invariant subspaces for Bose Hamiltonians. We consider systems whose
Hamiltonian can be expressed in terms of generators of $K_{i}$ (\ref{ki}).
Further, we (1) use the coherent state representation in which all Bose
operators become differential ones, (2) split the space of states of an
harmonic oscillator to even and odd states, (3) deform each of two pieces by
introducing, as a factor, additional unknown function which is different for
each piece, (4) demand that both these functions obey a coupled system of
differential equations that the action of $K_{i}$ convert each of basis
vectors into a linear combination of vectors of the same type, (5) select
coefficients, with which $K_{i}$ enter Hamiltonian $H$, to ensure the cut
off in the space of basis functions. In a sense, we introduce a kind of an
additional degree of freedom - effective ''spin'' $s$. Then kinds of
subspaces considered in our paper can be assigned the values $s=0$ (Sec.II), 
$1/2$ (Sec. III), $1$ (Sec. VII), etc. However, we want to stress that this
''spin'', in contrast to matrix QES models \cite{yve} - \cite{ren}, does not
appear in Hamiltonian and serves to describe the structure of solutions only.

It is worth stressing that it is just $H$ itself is quasi-exactly solvable,
whereas generators $K_{i}$ themselves, with the help of which $H$ is built
up, are, in general, not. This is in a sharp contrast with ''usual'' QES
Hamiltonians in quantum mechanics. Let us remind ourselves that in the
latter case $H$ can be expressed in terms of operators of an effective spin $%
S_{i}$ that realize $sl(2,R)$ algebra, each of them possessing
finite-dimensional subspace.

The essential ingredient of our approach is using the coherent state
representation. Formally, formulas (\ref{az}) look like those in the
coordinate-momentum representation. However, for our Hamiltonian one should
check carefully the normalizability of solutions that implies integration
over whole complex plane in a scalar product and impose constraints on
admissible values of parameters. Another constraints stem from the demand of
analyticity.

We want to point out that non-Hermitian QES operators can also be of
interest. They may be used, as auxiliary quantities, in physical
applications for finding spectra of Hermite Hamiltonians. For instance, they
may appear in mappings like $HL=LH^{\prime }$, where $H$ is Hermitian.
Knowing the spectrum of $H^{\prime }$ or its part due to its quasi-exact
solvability, one can restore a part of spectrum of a physical Hamiltonian $H.
$ Apart from this, the approach considered in the present article opens a
way to the search and classification of linear differential operators with
different invariant subspaces. This would enable one to generalize or extend
the results obtained for such subspaces with a basis of monomials \cite{tjmp}%
, \cite{trus}. In particular, for the case (\ref{bes}) we obtained solutions
in the form of combinations of Bessel functions and monomials.

In this paper we restricted ourselves to one-particle systems but the
suggested approach is obviously extendable to many-particle Hamiltonians. It
also enables one to generate new QES Bose Hamiltonians by choosing another
kinds of functions $A$, $B$, $C$, $D$ in eqs. (\ref{u}), (\ref{v}).

The suggested approach can be useful in the problems of solid state physics
when interaction between phonons is essential, quantum optics, theory of
molecules, etc. In this context, especially important is the fact that our
approach is extendable to many-particle systems. Concrete elaboration and
applications of the obtained results deserve special treatment.

\section{Acknowledgment}

O. Z. thanks for hospitality Claus Kiefer and Department of Physics of
Freiburg University, where the part of this work has been performed, and
gratefully acknowledges financial support from DAAD.





%
%

%
%

\end{document}